\documentclass[11pt,fleqn]{article}

\usepackage{amsmath,amsfonts,amssymb,latexsym,cite}
\usepackage{graphicx}

\def\noi{\noindent}

\newcommand{\Title}[1]{\noi {{\Large\bf #1}}\\[1ex]}

\def\Aunames#1{\noi{\bf #1}}
\def\au#1{${}^{#1}$}
\def\Addresses#1{\medskip\noi \protect
	\begin{description}\itemsep -3pt {\it #1} \end{description}}
\def\adr#1#2{\item[${}^{#1}$]{\it #2}}

\newcommand{\Abstract}[1]{\vskip 2mm \begin{center}
        \parbox{16.4cm}{\small\noi #1} \end{center}\medskip}

\def\email#1#2{\footnotetext[#1]{e-mail: #2}\addtocounter{footnote}{1}}

\def\nq{\hspace*{-1em}}
\def\nqq{\hspace*{-2em}}

\def\cm{\hspace*{1cm}}
\def\inch{\hspace*{1in}}

\def\Jl#1#2{#1 {\bf #2},\ }

\def\ApJ#1 {\Jl{Astroph. J.}{#1}}
\def\CQG#1 {\Jl{Class. Quantum Grav.}{#1}}
\def\DAN#1 {\Jl{Dokl. AN SSSR}{#1}}
\def\GC#1 {\Jl{Grav. Cosmol.}{#1}}
\def\GRG#1 {\Jl{Gen. Rel. Grav.}{#1}}
\def\JETF#1 {\Jl{Zh. Eksp. Teor. Fiz.}{#1}}
\def\JETP#1 {\Jl{Sov. Phys. JETP}{#1}}
\def\JHEP#1 {\Jl{JHEP}{#1}}
\def\JMP#1 {\Jl{J. Math. Phys.}{#1}}
\def\NPB#1 {\Jl{Nucl. Phys. B}{#1}}
\def\NP#1 {\Jl{Nucl. Phys.}{#1}}
\def\PLA#1 {\Jl{Phys. Lett. A}{#1}}
\def\PLB#1 {\Jl{Phys. Lett. B}{#1}}
\def\PRD#1 {\Jl{Phys. Rev. D}{#1}}
\def\PRL#1 {\Jl{Phys. Rev. Lett.}{#1}}

\def\lal{&&\nqq {}}
\def\eq{Eq.\,}

\def\beq{\begin{equation}}
\def\eeq{\end{equation}}
\def\bear{\begin{eqnarray}}
\def\bearr{\begin{eqnarray} \lal}
\def\ear{\end{eqnarray}}
\def\earn{\nonumber \end{eqnarray}}

\def\nnn{\nonumber\\ \lal }

\def\e{{\,\rm e}}

\def\diag{\mathop{\rm diag}\nolimits}

\def\eps{\varepsilon}

\def\aver#1{\left\langle \, #1 \, \right\rangle \mathstrut}

\def\mn{_{\mu\nu}}
\def\MN{^{\mu\nu}}
\def\mN{_\mu^\nu}
\def\nM{_\nu^\mu}
\def\cK{{\cal K}}

\def\kappa{\varkappa}

\def\R{{\mathbb R}}
\def\S{{\mathbb S}}

\def\rf{\eqref}
\def\eqn{\eq\eqref}

\usepackage{color}

\def\bh{black hole}
\def\bhs{black holes}
\def\Swz{Schwarz\-schild}

\begin{document}
\thispagestyle{empty}
\twocolumn[

\Title{The Schwarzschild singularity: a semiclassical bounce?}

\Aunames{S. V. Bolokhov\au{a,1}, K. A. Bronnikov\au{a,b,c,2}, 
		and M. V. Skvortsova\au{a,3}} 

\Addresses{
\adr a {\small Peoples' Friendship University of Russia (RUDN University), 
             ul. Miklukho-Maklaya 6, Moscow 117198, Russia}
\adr b {\small Center for Gravitation and Fundamental Metrology, VNIIMS,
             Ozyornaya ul. 46, Moscow 119361, Russia}
\adr c {\small National Research Nuclear University ``MEPhI''
                    (Moscow Engineering Physics Institute), Moscow, Russia}
	}

\Abstract
  {We discuss the opportunity that the singularity inside a Schwarzschild black hole could be replaced
    by a regular bounce, described as a regular minimum of the spherical radius (instead of zero) 
    and a regular maximum of the longitudinal scale (instead of infinity) in the corresponding 
    Kantowski-Sachs metric. Such a metric in a vicinity of the bounce is shown to be a solution 
    to the Einstein equations with the stress-energy tensor representing vacuum polarization of
    quantum matter fields, described by a combination of curvature-quadratic terms in the effective 
    action. The indefinite parameters of the model can be chosen in such a way that it remains a few
    orders of magnitude apart from the Planck scale (say, on the GUT scale), that is, in a 
    semiclassical regime. 
  }

]       
\email 1 {boloh@rambler.ru}
\email 2 {kb20@yandex.ru}
\email 3 {milenas577@mail.ru}

\section{Introduction}

  The existence of cosmological and black hole singularities is well known as a natural,
  though undesirable feature of general relativity (GR) and many alternative theories of 
  gravity.  It still appears that most of the researchers do not believe that infinite values
  of curvature invariants and/or matter densities and temperatures, inherent to such
  singularities, really exist in nature. It seems to be much more plausible that a modified
  theory of gravity must replace GR at large curvatures (or at high energies and 
  the corresponding small length and time scales), and that such a modification should 
  be related to quantum phenomena.  
  
  In the extremely numerous attempts to avoid singularities in the description of nature, 
  three basic trends may be singled out:\footnote
  	{We here do not even try to give a full list of references which would be enormously 
  	long; instead, we only mention some known reviews and some of the papers that we 
  	discussed in the course of this study.}   
\begin{description}   
\item[(a)] 
  Various models of quantum gravity, which should be treated as tentative ones 
  since a consistent and generally accepted theory of quantum gravity is so far lacking
  \cite{QG1, QG2, QG3, QG4}; 
\item[(b)] 
  Models of semiclassical gravity, treating gravity itself as a classical field and using
  the equations of GR or another classical theory of gravity to describe the geometry, 
  but including certain averages of quantum matter fields as sources of gravity
  \cite{SCG1, SCG1a, SCG2, SCG3, garr-07, hiscock-97, corda-11, 
  bardeen-14, kaw-13, kaw-17, mar-17};
\item[(c)]
  Inclusion of various classical sources of gravity, violating the usual energy conditions, 
  such as, for example, phantom scalar fields in GR, effective stress-energy tensors
  originating from additional  geometric quantities like torsion, or those borrowed from 
  extra space-time dimensions
  \cite{vis-book, BR-book, lobo-rev, kb-fab-02, kb-NED-01, bu-06, we-12, wh-16}. 
\end{description}

 One can notice a deep similarity between the singularity problems in the 
 Big Bang cosmology and those occurring in \bh\ interiors, at least those like
 the \Swz\ singularity, located in a ``T-region'' of the \bh, where the metric describes
 a special case of Kantowski-Sachs anisotropic cosmology. It is therefore not surprising 
 that  the same tools are used to regularize the cosmological and \bh\ singularities. 
    
 Models of quantum gravity \cite{QG1,QG2,QG3,QG4} 
 (or, more precisely, their effective representations written in the language of classical
 geometry) provide nonsingular models both in \bh\ physics and cosmology, but their
 general shortcoming is that they stop a collapse at curvatures and densities on the 
 Planck scale or very close to it. Since quantum matter fields demonstrate quantum 
 properties already at the atomic or even macroscopic scales (recall, e.g., lasers or the
 Casimir effect), there can be a hope that a cosmological collapse or \bh\ singularities 
 may be prevented at scales not so far from the conventional ones as is the Planck scale.
 This looks more attractive from an observational viewpoint as well as from the positions
 of theory since the corresponding results would look, at least to-date, more confident
 than those of quantum gravity. As to the third trend (c), although it has brought about
 a great number of studies of great interest such as models of cosmological bounce, 
 regular \bhs, wormholes, etc.
 (see, e.g., \cite{vis-book, BR-book, lobo-rev, kb-fab-02, kb-NED-01, bu-06, we-12, wh-16}),
 one should admit that, on one hand, their necessary ``exotic'' components require 
 conjectures not yet confirmed by the experiment, and, on the other, their possible existence 
 is usually treated as a phenomenological description of some underlying quantum effects.
 
 In this paper, we try to take into account the effects of quantum fields at approach to the
 \Swz\ singularity ($r=0$, often incorrectly called the central singularity whereas it is actually
 cosmological in nature).

 We consider a toy model in which the interior of the Schwarzschild space-time is
 studied within the scope of semiclassical approach for a possible bounce in its interior
 instead of the singularity. We try to take into account that any space-time always
 contains quantum oscillations of all physical fields, but here we do not make any 
 assumptions on their particular composition, restricting ourselves to pure vacuum
 polarization effects supposed to support a sought-for bouncing solution. In this simplified
 statement of the problem it turns out that there is a wide choice of the free parameters
 of the model providing realization of such a scenario.
 
 Unlike the studies trying to take into account the effects of Hawking radiation inside
 black holes, e.g., \cite{piran-94, corda-11, bardeen-14, kaw-13, kaw-17},
 we only discuss vacuum effects close to a would-be singularity. Such a description can 
 probably be relevant to sufficiently large \bhs, say, of stellar mass or those in galactic nuclei, 
 for which  the influence of Hawking radiation may be neglected. That is, it may be a possible 
 answer to the following question: if a body (a particle, a planet, a spacecraft) falls into a large \bh,
 which geometry will it meet there?

\section{The metric at bounce}

   The interior (T-region) of a generic spherically symmetric  \bh\
   space-time in the appropriate coordinate system can be described by the metric
\beq       \label{ds0}
    	 ds^2= d\tau^2 -\e^{2\gamma(\tau)}dx^2-\e^{2\beta(\tau)}d\Omega^2,
\eeq
   where $\tau$ is the ``cosmologiical'' time coordinate, and $x$ is a spatial coordinate
   that appears beyond the horizon replacing the time coordinate of the static region,
   $d\Omega^2$ is the metric on a unit sphere $\S^2$, while
   $\beta$ and $\gamma$ are smooth functions of time. This metric describes a 
   homogenious anisotropic cosmology of Kantowski---Sachs type, 
   with spatial section topology $\R \times \S^2$. 
   
   Let us assume that sufficiently far from a singularity this space-time corresponds to 
   the classical Schwarzschild solution, which in its T-region ($r < 2m$) has the form  
   ($m = GM$, where $M$ is the \bh\ mass, and we are using the units convention 
   $\hbar=c=1$).
\beq     \label{Swz}  \nq
	  ds^2 = \Big(\frac {2m}{T}-1\Big)^{-1} dT^2  - \Big(\frac {2m}{T}-1\Big)dx^2 
	  	-T^2 d\Omega^2,
\eeq   
  where, as compared to the usual expression, we have re-denoted $r \to T$ because in 
  the T-region the coordinate $r$ is temporal. At small $T$, substituting 
   $\sqrt{T/(2m)} dT = d\tau$, we adjust the asymptotic form of the metric at small $T$ 
   to the form \rf{ds0}:
\bearr                     \label {ds1}
	 ds^2 = d\tau^2 - \Big(\frac 43 m\Big)^{2/3} \tau^{-2/3} dx^2 
\nnn \inch	 
	 					- \Big(\frac 92 m\Big)^{2/3} \tau^{4/3}  d\Omega^2,   
\ear      
  valid at $\tau/m \ll 1$. As $\tau \to 0$, the scale along the $x$ axis infinitely stretches 
  while the coordinate spheres shrink to zero.      
   
  Our basic assumption in this study will be that the effect of quantum fields do not allow 
  the limit $r = \e^\beta \to 0$ but provide, instead, a regular minimum at some value $r=a > 0$, 
  while the scale along the $x$ axis has a regular maximum at $\tau=0$. Then at sufficiently 
  small $\tau$, in accord with \rf{Swz} and \rf{ds1}, the metric acquires the form 
\bearr 				\label{ds2}
         ds^2\Bigr|_{\rm bounce} \simeq d\tau^2 - \frac{2m}{a}(1-c^2\tau^2)dx^2
\nnn \inch \cm        
         			- a^2(1+b^2\tau^2) d\Omega^2
\ear
   with some positive constants $a, b, c$ of appropriate dimensions.\footnote
   	{Please do not confuse this $c$ with the speed of light.}

   Under this assumption, let us also assume symmetry of the metric 
   with respect to the bouncing time $\tau=0$ and, using the form \rf{ds0} of the metric,
   represent the functions $\beta(\tau)$ and $\gamma(\tau)$ as Taylor series with 
   even powers and constants $\beta_i, \gamma_i\; (i=0,2,4,6,...)$:  
\bearr \label{beta-gamma}
	\beta(\tau)=\beta_0+\frac{1}{2}\beta_2\tau^2
		+\frac{1}{24}\beta_4\tau^4+\frac{1}{720}\beta_6\tau^6+...
\nnn
	\gamma(\tau)=\gamma_0+\frac{1}{2}\gamma_2\tau^2
		+\frac{1}{24}\gamma_4\tau^4+\frac{1}{720}\gamma_6\tau^6+...,
\ear
  so that, in particular, according to \rf{ds2},
\bearr                                      \label{abc}  
	   	a=\e^{\beta_0}, \qquad   	2m/a  = \e^{2\gamma_0}, 
\nnn    		
    		2 b^2 = \beta_2/\beta_0,	\qquad 2 c^2 = - \gamma_2/\gamma_0.
\ear

  Now, in the semiclassical approach, we will seek the appropriate solution to the Einstein 
  equations 
\beq \label{EE}
		G\mN=-\varkappa \aver {T\mN}, \qquad \kappa = 8\pi G,
\eeq
   with the renormalized quantum stress-energy tensor $\aver{ T\mN}$, containing a contribution
   from vacuum polarization. 

   The nonzero components of the Einstein tensor for the metric \rf{ds0} are
\bearr          \label{Gmn}
	-G^0_0=\dot\beta(\dot\beta+2\dot\gamma)+\e^{-2\beta},
\nnn
	-G^1_1=2\ddot\beta+3\dot\beta^2+\e^{-2\beta},
\nnn
	- G^2_2  = - G^3_3 = \ddot\gamma+\ddot\beta
		+ \dot\gamma^2+\dot\beta^2+\dot\beta\dot\gamma.
\ear
     Using the Taylor expansion \eqref{beta-gamma}, one can write these components 
     explicitly up to $O(\tau^2)$ as follows:
\bearr
\nq		-G^0_0=\frac{1}{a^2}\left(1-\frac{\beta_2}
			{2\beta_0}\tau^2\right)+\beta_2(\beta_2+2\gamma_2)\tau^2,
\nnn
\nq		-G^1_1=\frac{1}{a^2}\left(1-\frac{\beta_2}
			{2\beta_0}\tau^2\right)+2\beta_2+\beta_4\tau^2+3\beta_2^2\tau^2,
\nnn
\nq		-G^2_2 = \beta_2\!+\!\gamma_2\!+\!\frac{1}{2}\!(\beta_4\!+\!\gamma_4)\tau^2\!
			+\!(\beta_2^2\!+\!\gamma_2^2\!+\!\beta_2\gamma_2)\tau^2,
\nnn{}
\ear
  while the r.h.s. of  \rf{EE} will be discussed in the next section.
   
\section{Stress-energy tensor}

   Following numerous papers on quantum field theory in curved spaces, in particular, the books 
   \cite{gmm, birrell}, we can present  the renormalized vacuum stress-energy tensor
   $T\nM$ in terms of a linear combination of certain geometric quantities ${}^{(i)}H\mn$ 
   ($i=1,2$) (obtainable by variation of actions quadratic in the Ricci tensor and scalar), 
   with phenomenological constants $N_1, N_2$, and two more contributions,
   ${}^{(c)}\!H\nM$ and $P\nM$:
\beq 					\label{SET}
	\aver{T\nM} = N_1 {}^{(1)}\!H\nM + N_2 {}^{(2)}\!H\nM + {}^{(c)}\!H\nM + P\nM
\eeq
   where
\bearr                                           \label{HH}
		 {}^{(1)}\!H\nM \equiv 2R R\nM-\frac{1}{2}\delta\nM R^2
		 	+2\delta\nM \Box R-2\nabla_\nu\nabla^\mu R,
\nnn
	{}^{(2)}\!H\nM \equiv
	-2\nabla_\alpha\nabla_\nu R^{\alpha\mu}+\Box R\nM+\frac{1}{2}\delta\nM\Box R
\nnn
	\cm\; +2R^{\mu\alpha} R_{\alpha\nu}-\frac{1}{2}\delta\nM R^{\alpha\beta}R_{\alpha\beta},
\ear
   and $\Box = g^{\mu\nu}\nabla_\mu \nabla_\nu$. The tensor ${}^{(c)}\!H\nM$ is a local
   contibution depending on space-time topology or special boundary conditions (the Casimir effect 
   \cite{Milton, Elizalde}), while $P\nM$ depends on the choice of the quantum state and 
   describes, in particular, particle production due to a nonstationary nature of the metric. The tensor 
   $P\mN$ is nonlocal in the sense that it is not a function of a point in space-time but depends on 
   the whole previous history. Its calculation is rather complicated and, moreover, depends on 
   additional assumptions on the quantum states of the constituent fields. We will not consider it
   in the present paper, assuming that its contribution can be insignificant at least for some 
   admissible choices of quantum states. 

  The components of the tensors ${}^{(i)}\!H^\mu_\nu$ (which turns out to be diagonal) are 
   easily  found using the ansatz \rf{ds0} and the Taylor series \rf{beta-gamma}. At $\tau =0$,
   that is, at bounce (higher orders in $\tau$ will not be necessary in our calculations) they are
\bearr      \label{Hmn}
            	{}^{(1)}\!H^0_0 = -\frac{2}{a^4}+8\beta_2^2+8\beta_2\gamma_2+2\gamma_2^2,
\nnn
          	{}^{(1)}\!H^1_1 = -\frac{2}{a^4} \!
         	-32\beta_2^2 - 16\beta_2\gamma_2 - 6\gamma_2^2\! - 8\beta_4 - 4\gamma_4,
\nnn
           	{}^{(1)}\!H^2_2 = 
         	  \frac{2}{a^4}+\frac{12\beta_2}{a^2}-24\beta_2^2 - 20\beta_2\gamma_2-10\gamma_2^2
\nnn \inch
         	 - 8\beta_4 - 4\gamma_4,
\nnn
		   	{}^{(2)}\!H^0_0 = -\frac{1}{a^4}+3\beta_2^2+2\beta_2\gamma_2+\gamma_2^2,
\nnn
		{}^{(2)}\!H^1_1 = 
		-\frac{1}{a^4}-9\beta_2^2 - 6\beta_2\gamma_2-3\gamma_2^2 - 2\beta_4 - 2\gamma_4,
\nnn
	      	{}^{(2)}\!H^2_2 = 
			    \frac{1}{a^4} + \frac{4\beta_2}{a^2}-9\beta_2^2 - 6\beta_2\gamma_2 -3\gamma_2^2
\nnn \inch
	    		 - 3\beta_4 - \gamma_4.
\ear
   
   The numerical coefficients $N_1$ and $N_2$ in \rf{SET} are unknown and should be determined
    from experiment or observations. One can qualitatively estimate the order of magnitude for these
    coefficients recalling that they actually appear as coefficients at curvature-squared terms 
    in higher-derivative gravity models with action of the form 
    $S\sim \int d^4x\sqrt{-g}(R/(2\kappa) + N_1 R^2 + N_2 R^2_{\mu\nu}+...)$
    since the tensors ${}^{(1,2)}H\mn$ result from variation of terms with $R^2$ and 
    $R^2\mn \equiv R\mn R\MN$ with respect to the metric \cite{gmm, birrell}.
    
    The current empirical upper bound for these parameters is $N_{1,2}   \lesssim 10^{60}$ 
    (see \cite{Giacchini}). This restriction follows from observations performed at very 
    small curvatures, therefore possible effects of terms $\sim R^2$ are hardly 
    noticeable. The estimates of $N_{1,2}$ may be different if such theories of gravity 
    are applied to the early (inflationary) Universe where curvatures are much larger,
    for example, $N_1 \sim 10^{10}$ \cite{Starob0, Starob,Odin}. In any case, for our purposes 
    we can feel more or less free in the choice of these parameters.  
   
   As to the Casimir contribution, there are reasons to believe that it should be small as compared 
   to that of ${}^{(i)}H\mn$. Consider, for example, the static case of the metric \rf{ds0} with 
   $\e^\beta = r =a$, actually describing what can be called an infinitely long wormhole throat.   
   A calculation performed in \cite{butcher} for a massless conformally coupled scalar field
   gives for this geometry
\bearr     \label{Cas}
	   {}^{(c)}\!H\nM = 
	   \frac{1}{2880\pi^2a^4}\Big[ 2\diag(-1,-1,1,1) \ln(a/a_0) 
\nnn \inch	   
	   + \diag(0, 0, -1, -1)\Big], 	   
\ear
  where $a_0$ is a fixed length that can only be determined by experiment. The quantity \rf{Cas}
  corresponds to a single massless scalar, while the total Casimir contribution must 
  include all fields with different spins and masses.
  
  On the other hand, for the same geometry, 
\beq
             {}^{(1)}H\mN = 2{}^{(2)}H\mN = \frac{2}{a^4} \diag (-1, -1, 1, 1).
\eeq    
  Thus, if $N_1$ and/or $N_2$ are of the order of unity or larger (as we shall finally need),
  the tensors ${}^{(i)}H\mn$ will evidently much stronger contribute to $\aver{T\mN}$ than
  ${}^{(c)}H\mn$, unless the uncertain quantity $a_0$ in \rf{Cas} is unnaturally high, or the total 
  number of fields is so great as to overcome the denominator of about $10^4$. 
  In our further estimates we will assume that ${}^{(c)}\!H\nM$ can also be neglected 
  in our more general geometry and thus take into account only ${}^{(i)}H\mN$.
  
\section{Semiclassical bounce}

  Consider the Einstein equations \eqref{EE} with the stress-energy tensor 
  \rf{SET}, taking there into account only the first two terms, and let us find out whether there 
  are solutions consistent with our assumption on the bouncing metric \rf{ds2}, and if yes,
  what are restrictions on the free parameters of the model providing their semiclassical scale.       
  
  To do that, we should express $G\nM$ and ${}^{(i)}H\mn$ in terms of the Taylor
   series~\eqref{beta-gamma} and equate coefficients at equal powers of $\tau$ on the two 
   sides of resulting equations. For convenience, we introduce the dimensionless parameters
\bearr                       \label{ABC}
   	A = \kappa a^{-2}, \quad 
   	B_2 = \kappa\beta_2, \quad 
   	C_2 = \kappa\gamma_2, 
\nnn   	
   	B_4=  \kappa^2\beta_4,\quad 
   	C_4 = \kappa^2\gamma_4, \ \ {\rm etc.}
\ear   
   Since $\kappa \approx l_{\rm pl}^2$ (the Planck length squared), it is clear that our system 
   will remain on the semiclassical scale as long as all parameters \rf{ABC} are much smaller 
   than unity. This means, in particular, that the minimum radius $r = a$, reached  at bounce,
   will be much larger than the Planck length. Other parameters are values of the 
   derivatives $\ddot \beta, \ddot\gamma$, etc. at bounce. 
      
   It turns out that in the approximation used it is sufficient to check only the order $O(1)$
   in the ${0 \choose 0}$ component of the equations. This yields
\bearr                                                                   \label{00-0}
   	A = N_1 [-2 A^2+2(2B_2+C_2)^2]
\nnn  \cm 
			 +N_2 [-A^2 +(B_2+C_2)^2+2B_2^2].
\ear
  Other equations will only express the constants $B_4, C_4$, etc. in terms of $A, B_2, C_2$.
  So we have only one equation for the latter three parameters and also $N_1, N_2$ and 
  have a large space of possible solutions.
  
  According to our purposes, we assume that $a$ is much larger than the Planck length 
  $ l_{\rm pl} \sim \sqrt{\kappa}$, which means that $A \ll 1$, or $A = O(\eps)$, $\eps$ 
  being a small parameter. Let us also make a natural assumption that  
  $B_2, C_2 = O(\eps)$, that is, $\ddot\beta$ and $\ddot\gamma$ are of the same 
  order of magnitude as $1/a^2$. Then, since the r.h.s. of \eqn{00-0} is $O(\eps^2)$ while the 
  l.h.s. is $O(\eps)$, to maintain the equality we must have $N_1$ and/or $N_2$ being 
  large enough, of order $O(1/\eps)$. 
  
  From the remaining Einstein equations ${1 \choose 1}$ and ${2 \choose 2}$ at $\tau =0$ 
   it then follows that $B_4$ and $C_4$ are of order $O(\eps^2)$ (see \rf{HH}), 
  so that the 4th order derivatives of $\beta$ and $\gamma$ are of a correct order of 
  smallness relative to the Planck scale (see \rf{ABC}). Similar inferences follow
  for $B_6, C_6$, etc. if we consider the equations in the order $O(\tau^2)$, and so on.
  One can also verify that the curvature invariants $R$, $R\mn^2$ and 
  $\cK\equiv R_{\mu\nu\rho\sigma}R^{\mu\nu\rho\sigma}$ are small at bounce ($\tau=0$)
  as compared to the Planck scale: 
\bearr \label{inv}   
    	R=  \frac{2}{a^2} + 4\beta_2 + 2\gamma_2 = O\Big(\frac{\eps}{\kappa}\Big),
\nnn   
        R\mn^2\!=\frac{2}{a^4}+\frac{4\beta_2}{a^2}+6 \beta_2^2 
                        +4 \beta_2\gamma_2+2\gamma_2^2=O\Big(\frac{\eps^2}{\kappa^2}\Big),
\nnn   
	\cK=\frac{4}{a^4}+8\beta_2^2+4\gamma_2^2=O\Big(\frac{\eps^2}{\kappa^2}\Big).
\ear

   Let us illustrate the situation with a numerical example. Assume $N_1 =0$, 
   $N_2= 10^{10}$,  and $A = 10^{-10}$ (hence a minimal radius $a$ of $10^5$ Planck lengths).
   Furthermore, since by construction  (see \rf{abc} and \rf{ABC}) $B_2 > 0$ and $C_2 <0$, 
   we can safely assume $B_2 + C_2 =0$.
   Under these assumptions, from \eqn{00-0} we find
\[
		B_2 = - C_2 = 10^{-10}.
\]            
  Substituting this into the ${1 \choose 1}$ and ${2 \choose 2}$ components of the 
  Einstein equations at $\tau =0$, using the expressions \rf{Gmn} and \rf{Hmn}, we obtain 
  the values of $B_4$ and $C_4$:
\[
  		B_4 = 3.5 \times 10^{-20}, \qquad C_4 = - 8.5 \times 10^{-20}.
\]
  From equations of order $O(\tau^2)$ one can then find $B_6, C_6$, and so on.

  It is known that at a regular minimum of the spherical radius $\e^\beta = r$, be it a 
  wormhole throat in an R-region or a bounce in a T-region, the stress-energy-tensor 
  must satisfy the requirement  $T^0_0 - T^1_1 < 0$ thus violating the Null Energy Condition, 
  see, e.g.,  \cite{BR-book, BKor-15}. In our case, since we suppose a bounce at $\tau =0$,
  the inequality $T^0_0 - T^1_1 < 0$ automatically holds as long as we use 
  the metric \rf{ds2}.  

\section{Concluding remarks}

  We can conclude that under our assumptions a semiclassical bounce instead of a \Swz\ 
  singularity is possible and can occur at semiclassical scales, without need for quantum 
  gravity effects. 

    As already mentioned, this result can tell us which kind of geometry may be seen by an
    observer falling into a sufficiently large \bh\ for which Hawking radiation is negligible due 
    to its extremely low temperature. Though, for a more complete and convincing answer 
    to the same question it would be necessary to take into consideration nonlocal effects 
    (above all, particle creation) depending on the choice of quantum states of different
    physical fields, also making clear which quantum states can be physically relevant.
   
    Of even greater interest can be similar studies for charge and rotating \bhs\ where the 
   singularities are not so simple as in the \Swz\ case. We hope to address these problems 
   in our future studies.   
         
    There seems to appear an attractive opportunity that the terms in the effective action 
    quadratic in the Ricci tensor, with $N_{1,2} \sim 10^{10}$ or so (that is, on the GUT 
    rather than Planck scale), being able to drive inflation in the early Universe 
    \cite{Starob0, Starob, Odin}, are also able to prevent \bh\ singularities 
    but are quite unnoticeable under usual physical conditions. Such a mechanism may be
    universal, as possibly indicated by the absence of the \bh\ mass in our equations for the 
    near-bounce metric. Avoidance of \bh\ singularities due to semiclassical effects was 
    obtained under other assumptions in \cite{SCG2,corda-11,bardeen-14,kaw-13,kaw-17}; 
    however, some other studies indicate that semiclassical effects may even enhance a singularity 
    (e.g., \cite{piran-94,hiscock-97}), so the results evidently strongly depend on the assumptions 
    made.

\subsection*{Acknowledgments}

  This publication has been prepared with the support the RUDN University Program 5-100
  and by RFBR grant 16-02-00602. The work of KB was also partly performed within the
  framework of the Center FRPP supported by MEPhI Academic Excellence Project 
  (contract No. 02.a03. 21.0005, 27.08.2013).


\end{document}